\documentclass{article}
\usepackage{spconf,amsmath,graphicx}
\usepackage{xcolor}
\usepackage{colortbl}
\usepackage{multirow}
\usepackage{booktabs}
\usepackage{url}

\usepackage[numbers,sort&compress]{natbib}
\usepackage[textsize=tiny,textwidth=0.7in, disable]{todonotes}

\title{ Spoofing-Aware Speaker Verification via Wavelet Prompt Tuning and Multi-Model Ensembles}

\name{Aref Farhadipour, Ming Jin, Valeriia Vyshnevetska, Xiyang Li, Elisa Pellegrino, Srikanth Madikeri}
\address{Department of Computational Linguistics, University of Zurich, Switzerland 
\\ \{aref.farhadipour, srikanth.madikeriraghunathan\}@uzh.ch}

\begin{document}
%
\maketitle
\begin{abstract}
This paper describes the UZH-CL system submitted to the SASV section of the WildSpoof 2026 challenge. The challenge focuses on the integrated defense against generative spoofing attacks by requiring the simultaneous verification of speaker identity and audio authenticity.
 We proposed a cascaded Spoofing-Aware Speaker Verification framework that integrates a Wavelet Prompt-Tuned XLSR-AASIST countermeasure with a multi-model ensemble. The ASV component utilizes the ResNet34, ResNet293, and WavLM-ECAPA-TDNN architectures, with Z-score normalization followed by score averaging. Trained on VoxCeleb2 and SpoofCeleb, the system obtained a Macro a-DCF of 0.2017 and a SASV EER of 2.08\%. While the system achieved a 0.16\% EER in spoof detection on the in-domain data, results on unseen datasets, such as the ASVspoof5, highlight the critical challenge of cross-domain generalization. The source code is available\footnote{\url{https://github.com/areffarhadi/SASV}}.

\end{abstract}
\begin{keywords}
SASV, Wavelet Prompt Tuning, Deepfake Detection, Speaker verification, SSL.
\end{keywords}
\section{Introduction}
\label{sec:intro}

Traditional speaker verification systems have focused mainly on modeling individual voice characteristics to answer the question of identity \cite{farhadipour2024analysis}. However, the rapid advancement of generative AI has made these systems vulnerable to sophisticated spoofing attacks, such as text-to-speech and voice conversion. The Integrated spoofing and speaker verification task has emerged as a critical frontier for addressing these threats, requiring systems to simultaneously verify "who is speaking" and "whether the speech is authentic." Benchmarks such as NIST speaker recognition evaluation and WildSpoof challenges \cite{wu2025wildspoof} have been instrumental in pushing the field towards unified and secure biometric authentication. Building on previous work in robust speaker modeling \cite{farhadipour2025cl}, we introduce a cascaded architecture for the SASV task.

In this work, we propose a cascaded SASV architecture that sequentially integrates a countermeasure (CM) with a multi-model Automatic Speaker Verification (ASV) ensemble. For spoofing detection, we implemented the Wavelet prompt fine-tuned XLSR \cite{baevski2020wav2vec} model with the AASIST backend \cite{jung2022aasist}, which, leverages Wavelet Prompt Tuning (WPT) to capture multi-resolution spectral artifacts. For the ASV component, we used an ensemble of three architectures consisting of ResNet34, ResNet293, and WavLM-ECAPA-TDNN. By training our systems on VoxCeleb2 \cite{chung2018voxceleb2} and SpoofCeleb \cite{jung2025spoofceleb}, we demonstrate that the cascaded approach effectively filters spoofed audio while maintaining high precision in speaker recognition, even under challenging, unseen attacks in the evaluation set.

\section{Methodology}
\label{sec:method}

\subsection{CM: Wavelet Prompt-Tuned XLSR with AASIST Back-end}
The proposed spoof detection (SD) system employs a multi-stage architecture that leverages the synergy between large-scale self-supervised learning (SSL) and frequency-aware, parameter-efficient tuning. The system's front-end is built upon the pre-trained XLSR-53 model, which serves as a robust feature extractor for high-dimensional latent representations from raw audio waveforms. 

To mitigate the computational overhead of full-parameter fine-tuning while enhancing the model's sensitivity to synthetic speech artifacts, a WPT mechanism is integrated into the SSL backbone's transformer layers. These enhanced embeddings are subsequently fed into the AASIST back-end, which employs a dual-graph attention network (GAT) structure. 

The system is trained using a weighted cross-entropy loss to maximize the decision margin between bonafide and spoofed speech. Optimization is conducted using the Adam optimizer with a learning rate of 1e-4 and a cosine annealing schedule. The WPT reduces the total number of trainable parameters to around 1M, while the XLSR model contains 300M parameters.

The WPT-XLSR-AASIST model \cite{xie2025detect} was implemented with a configuration consisting of 6 learnable prompt tokens and 4 specialized wavelet prompt tokens to facilitate the extraction of multi-resolution features. The system was trained for 15 epochs with a batch size of 64.

\subsection{ASV: Fusion of SSL and Resnet Models }
For the ASV component, we employed a multi-model ensemble consisting of three high-performance architectures: ResNet34, ResNet293, and WavLM-ECAPA-TDNN \cite{chen2022wavlm, desplanques2020ecapa}. The ResNet34 and ResNet293 models follow the technical specifications outlined in \cite{lin2024voxblink2}. They utilize Attentive Statistics Pooling to aggregate frame-level features and are optimized using the \cite{deng2019arcface}. These models were trained on the VoxCeleb2 dataset to ensure robust speaker embeddings across diverse acoustic environments. 

The third system utilizes a WavLM-Large front-end integrated with an ECAPA-TDNN downstream head and finetuned on VoxCeleb2. 
To combine the outputs of these three distinct systems, we implemented a Z-score averaging strategy. The final verification score is then computed using a weighted sum of the normalized scores.

\subsection{SASV System}
The proposed SASV system is implemented as a cascade architecture that sequentially integrates the CM and ASV modules to provide comprehensive security against both spoofing and impersonation. In the first stage, the test utterance is analyzed by the CM to classify the audio as either bonafide or a spoofing attack.

The second stage involves the ASV module, which processes the test utterance alongside the reference enrollment utterance. For speakers with multiple enrollment utterances, the final score for each test utterance was calculated as the average of all corresponding enrollment scores. This block utilizes the multi-model ensemble and the Z-score averaging strategy. The final output is a joint decision: a sample is only accepted as a "Target" if it is both verified as bonafide by the CM and matched to the correct identity by the ASV. 

\section{Experimental Resuls}
\label{sec:setup}

As shown in Table \ref{tab:asv_results}, we evaluated three different architectures for the ASV task. Among individual models, ResNet293 achieved the highest performance with an EER of 2.97\% on the SpoofCeleb evaluation set. However, implementing the Z-Score averaging strategy improved system reliability, resulting in a fused EER of 2.25\%. For the Spoof Detection task, our proposed model achieved an EER of 0.16\% on the evaluation set, confirming its high sensitivity to synthetic artifacts.

The performance of the integrated cascaded system is detailed in Table \ref{tab:sasv_results}. The SASV EER, reached 2.08\% on the evaluation set. Notably, the a-DCF was 0.0375 on the evaluation data.
Finally, we report the performance of our final submission in Table \ref{tab:final_test_adcf}\footnote{\url{https://wildspoof.github.io/91_results/}}. Our system was tested in four challenging datasets: WildSpoof-TTS, SpoofCeleb, ASVspoofF5 and ASV 2022 while trained on SpoofCeleb. The system achieved a Macro a-DCF of 0.2017. 

Although the system demonstrated robust performance on SpoofCeleb, the results on ASVspoofF5 (0.40) and ASV 2022 (0.32) underscore the significant challenges of cross-domain generalization and the detection of unseen spoofing attacks. These findings suggest that system performance could be further strengthened by incorporating a broader range of training data across a wider variety of attack scenarios.

\begin{table}[h]
\centering
\caption{Performance of Individual sub-modules and Z-Score Fusion in each ASV and SD tasks in EER \%. }
\label{tab:asv_results}
\resizebox{0.38\textwidth}{!}{%
\begin{tabular}{llcc}
\toprule
\textbf{Task} & \textbf{Model} & \textbf{Dev} & \textbf{Eval} \\
\midrule
\multirow{4}{*}{ASV} & ResNet34        & 4.22 & 3.74 \\
                     & WavLM-ECAPA-TDNN           & 3.36 & 3.86 \\
                     & ResNet293       & 3.35 & 2.97 \\
\cmidrule(lr){2-4}
                     & \textbf{Z-Score Fusion}  & \textbf{2.48} & \textbf{2.25} \\
\midrule
 CM  & WPT-XLSR & 0.45 & 0.16 \\
\bottomrule
\end{tabular}%
}
\end{table}

\begin{table}[h]
\centering
\caption{Performance of the proposed cascade system in detail}
\label{tab:sasv_results}
\resizebox{0.45\textwidth}{!}{%
\begin{tabular}{lcccc}
\toprule
\textbf{Data} &  \textbf{SD (EER\%)} & \textbf{ASV (EER\%)} & \textbf{SASV (EER\%)} & \textbf{a-DCF} \\
\midrule
Dev       & 1.21 & 2.72 & 2.31 & 0.0469 \\
Eval      & 0.23 & 2.35 & 2.08 & 0.0375 \\
\bottomrule
\end{tabular}%
}
\end{table}

\begin{table}[h]
\centering
\caption{Performance of final submission of UZH-CL team, in a-DCF}
\label{tab:final_test_adcf}
\resizebox{0.47\textwidth}{!}{%
\begin{tabular}{lccccccc}
\toprule
\textbf{Team ID} & \textbf{WildSpoof-TTS} & \textbf{SpoofCeleb} & \textbf{ASVspoofF5} & \textbf{ASV 2022} & \textbf{Macro a-DCF}  \\
\midrule
T01 	& 0.0870 	& 0.0856 	& 0.0419 	& 0.2320 	& 0.0347\\
T02 	& 0.1329 	& 0.1611 	& 0.0269 	& 0.2543 	& 0.1670\\
T03 (ours) & 0.1924 & 0.0457 & 0.4088 & 0.3252 & \textbf{0.2017} \\
\bottomrule
\end{tabular}%
}
\end{table}

\section{Conclusion}
\label{sec:conclusion}

In this work, we explained our proposed SASV system for WildSpoof challenge. By combining the multi-resolution spectral awareness of Wavelet Prompt Tuning with a high-capacity ensemble of residual and self-supervised speaker verification models, we achieved a Macro a-DCF of 0.2017. Our findings indicate that Z-score fusion is an effective strategy for aligning diverse ASV architectures, while the WPT-XLSR-AASIST configuration offers a parameter-efficient yet powerful solution for spoofing detection. Overall, our results highlight that careful integration of spectral, residual, and self-supervised features can achieve competitive performance in SASV benchmarks, providing a promising direction for future research in robust speaker verification under spoofing attacks.


\bibliographystyle{IEEEbib}
\bibliography{refs}

\end{document}